\documentclass[fleqn, twocolumns, 10pt]{wlscirep}
\usepackage{doi}
\usepackage{longtable}
\usepackage[utf8]{inputenc}
\usepackage[T1]{fontenc}
\usepackage[none]{hyphenat}
\usepackage{hyperref}
\hypersetup{colorlinks=true, citecolor=blue, filecolor=blue, linkcolor=blue, urlcolor=blue}
\title{Computational Discovery of Inorganic Electrides from an Automated Screening}

\author[1,*]{Qiang Zhu}
\author [2]{Timofey Frolov}
\author [3]{Kamal Choudhary}

\affil[1]{Department of Physics and Astronomy, University of Nevada, Las Vegas, NV, 89154, USA}
\affil[2]{Lawrence Livermore National Laboratory, Livermore, CA, 94550, USA}
\affil[3]{Materials Science and Engineering Division, National Institute of Standards and Technology, Gaithersburg, MD, 20899, USA}
\affil[*]{Lead Contact: qiang.zhu@unlv.edu}

\date{\today}
\begin{abstract}
Electrides, with their excess electrons distributed in crystal cavities playing the role of anions, exhibit a variety of unique properties which make these materials desirable for many applications in catalysis, nonlinear optics and electron emission. While the first electride was discovered almost four decades ago, only few electride materials are known today, which limits our fundamental understanding as well as the practical use of these exotic materials. In this work, we propose an automated computational screening scheme to search for interstitial electrons and quantify their distributions inside the crystal cavities and energy space. Applying this scheme to all candidate materials in the Inorganic Crystal Structure Database (ICSD), we report the identification of 167 potential electride materials. This work significantly increases the current library of electride materials, which enables an in-depth investigation of their chemistry-structure-property relations and opens new avenues for future electride design.
\end{abstract}

\begin{document}

\flushbottom
\maketitle

\textbf{keywords:} Electride, high-throughput calculation, density functional theory, catalysis, topological materials

\section*{Introduction}

Electides represent a unique class of materials in which the excess electrons are trapped inside crystal cavities and serve as the anions \cite{Dye-Science-1990, dye2009electrides}. Localization of electrons provides the early examples of quantum confinement. These trapped electrons in an electride are usually loosely bound and form unique interstitial energy bands around Fermi level, which dominates the transport and magnetic properties. Engineering on these energy bands can be used to design new materials with low work function and high carrier mobility, which make electrides attractive for various materials applications. While the first crystalline organic electride was made by Dye and coworkers in 1983 \cite{Ellaboudy-1983-JACS}, the use of organic electrides for practical applications was limited by their thermal instability \cite{dye2009electrides}. A significant progress towards the materialization of electrides was made by Hosono and coworkers \cite{Matsuishi-Science-2003}. Starting from the mineral mayenite (12CaO$\cdot$7Al$_2$O$_3$), they synthesized the first room-temperature stable inorganic electride Ca$_6$Al$_7$O$_{16}$ (C12A7:2e$^-$) via oxygen-reducing processes. Since then C12A7:2e$^-$ has been used for ammonia synthesis \cite{Kitano-NChem-2012, Kuganathan-JACS-2014, Hayashi-JACS-2014} and as an electron-injection barrier material \cite{Hosono-PNAS-2017}. The discovery of C12A7:2e$^-$ has stimulated many new efforts to search for other inorganic electrides. Several new materials with improved functionality were developed in the recent years \cite{li2004theoretical, dye2005alkali, Lee-Nature-2013, Zhang-JPCL-2015, Zhang-QM-2017, Wang-JACS-2017, Lu-JACS-2016, Inoshita-PRX-2014, Tada-IC-2014, Zhang-PRX-2017, burton2018high}. 

Our fundamental understanding of electrides have been greatly expanded in the recent years, thanks to the growing number of the discovered electride materials. Early discovered organic electrides exhibited very different magnetic properties ranging from almost no magnetism to having a strength comparable to $k_\textrm{B}T$ at room temperature, depending on the geometry of the trapped electrons \cite{dye2009electrides}. Magnetic electrides were reported in inorganic materials as well \cite{Pickard-PRL-2011,inoshita2015ferromagnetic, Wang-JACS-2017}. More recently, it has been suggested that electrides are suitable for achieving various topological phases in condensed matter physics, such as Y$_2$C \cite{Huang-NanoL-2018, Hirayama-PRX-2018}, Sc$_2$C \cite{Hirayama-PRX-2018}, Sr$_2$Bi \cite{Hirayama-PRX-2018}, Ca$_3$Pb \cite{Ca3Pb-2018}, Cs$_3$O/Ba$_3$N \cite{Park-PRL-2018} and Rb$_3$O \cite{Zhu-PRM-2019}. Pressure has also proved to be an effective means to advance the electride research. For example under compression, many simple metals (Li \cite{Pickard-PRL-2009, Matsuoka-Nature-2009}, Na \cite{Ma-Nature-2009}, K \cite{Pickard-PRL-2011}), and compounds (NaHe$_2$\cite{Dong-NChem-2017}, Mg$_3$O$_2$ \cite{Zhu-PCCP-2013}) were found to adopt structures with valence electrons localized in the interstitial regions. These high pressure electrides (HPEs) were shown to exhibit a variety of intriguing physical phenomena, such as metal-semiconductor-insulator transition \cite{Ma-Nature-2009, Lv-PRL-2011, Pickard-PRL-2009, Matsuoka-Nature-2009} and superconductivity \cite{Shimizu-Nature-2002}. 

In the past, the identification of electrides required labour-intensive efforts which included raw materials synthesis and sample characterization. Considering the current size of materials data ($\sim$30000 after the removal of duplicates and structures with partial occupation) in the Inorganic Crystal Structure Database (ICSD) \cite{ICSD}, it is impossible to investigate each compound by trial-and-error experiments. The high-throughput (HT) computational materials design, based on a search for target materials from a large database containing necessary thermodynamic and electronic properties for all available materials, has become popular in materials science in recent years \cite{Curtarolo-2013}. To enable electrides discovery with HT strategies one can proceed by (1) screening the materials with excess electrons localized in the crystal cavities and (2) identifying the real space and energy distribution of the interstitial electrons. In this work, we address this challenge by developing a set of reliable yet computationally feasible descriptors to quantify the interstitial electrons in a multi-dimensional space. Applying these descriptors to the existing ICSD materials included in the open database of Materials Project \cite{MP-2013}, we report the identification of a subset of materials which possess interstitial electrons around Fermi level ($E_\textrm{F}$), 114 of which have not been considered in the previous literature. The total of 167 materials exhibits a diverse distribution in the chemical space and can serve as a guide for further in-depth studies.

\begin{figure}[ht]
\centering
\includegraphics[width=0.9\linewidth]{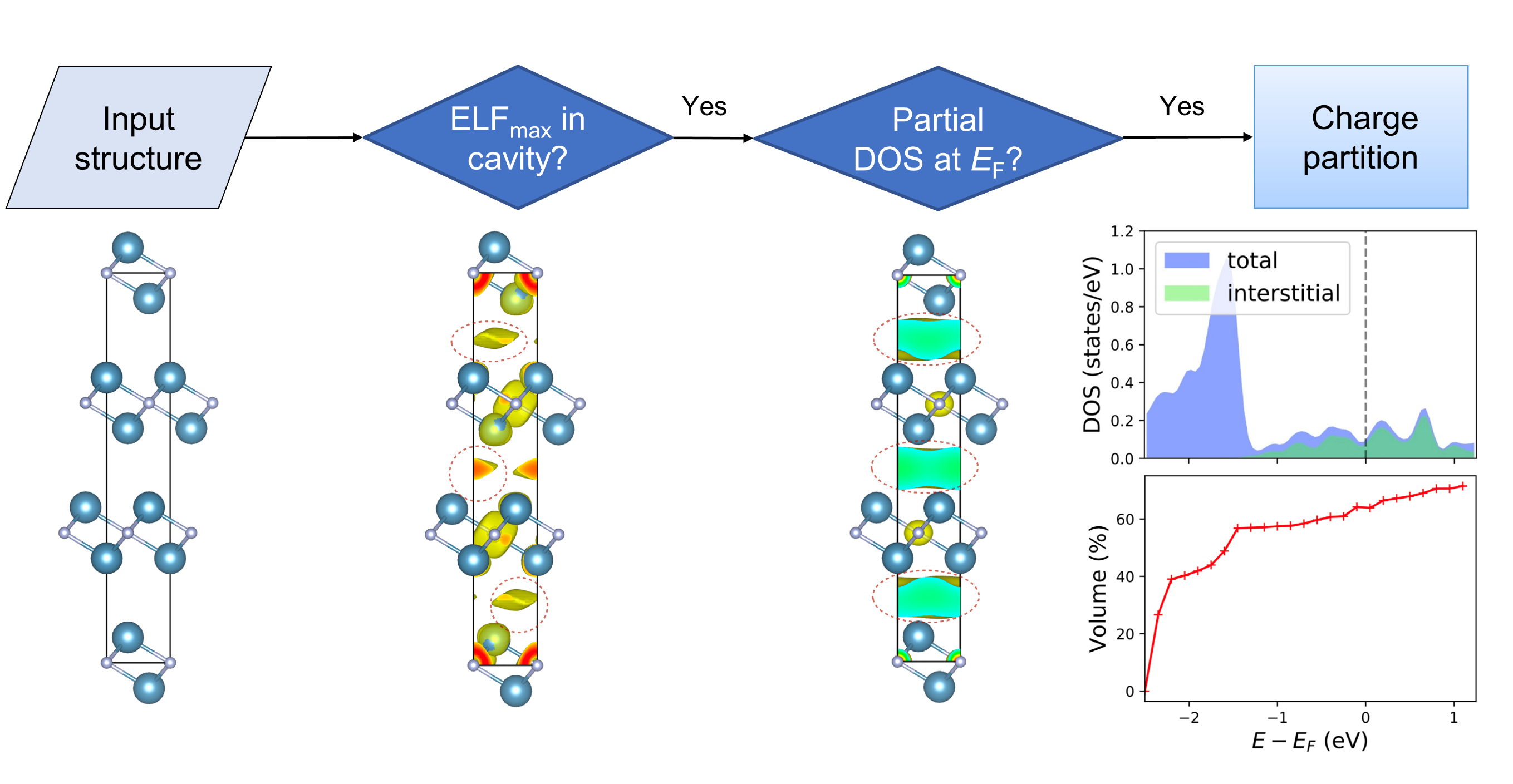}
\caption{\textbf{The computational scheme for inorganic electride identification.}} 
\label{fig:scheme}
\end{figure}

\section*{RESULTS AND DISCUSSION}
While in many cases materials can be simply classified by a single physical property (e.g., the band gap values can be used to distinguish between metal, semiconductor and insulator), a descriptor to distinguish electrides from other types of materials is not known at present. Several HT strategies for electride discovery proposed in the past included a search based on structural prototype \cite{Inoshita-PRX-2014,Tada-IC-2014}, identification of materials with strong electron localization in the interstitial space \cite{Zhang-PRX-2017} and the analysis of the partial density of states (PDOS) around $E_\textrm{F}$ \cite{burton2018high}. However, none of these approaches seem to provide a complete and accurate description of electrides. For instance, such strategies fail to identify the first experimentally established inorganic electride C12A7. More extensive theoretical work on characterizing the electrides have been conducted recently \cite{Li-Review-2016, dale-2018}. Based on the previous work and our own analysis performed on the known electride materials, we summarize three most important features of electrides, including (1) interstitial electrons; (2) ionic bonding and (3) floating around or adjacent to Fermi Level. While the existence of interstitial electrons is certainly the most important descriptor for an electride, we emphasize that the presence of the other two features is equally necessary. The anionic nature makes the excess electrons intrinsically different from (nearly) free electrons in simple metals. Furthermore, these excess electrons need to float around Fermi level, so that they can contribute to low work function and high mobility, which are important for practical applications. Accordingly, we designed a computational screening strategy which consists of multiple steps including the analysis of electron localization function (ELF) \cite{ELF}, band structure, partial charge density and PDOS as shown in Fig. \ref{fig:scheme}. First, we analyze the ELF, as the existence of interstitial ELF maximum (regardless of its magnitude) can serve as the first evidence for an electride \cite{Li-Review-2016, Zhang-PRX-2017}. Second, we calculate the partial charge density of this compound at an energy range within $E_\textrm{F}$ $\pm$ 0.05 eV (see details in the supplementary section 1). If there also exists charge maximum in the same interstitial region, we classify it as a possible electride and proceed to a more detailed analysis on the topology and energy distribution of the interstitial electrons. Following the Bader charge partition scheme \cite{Bader-1990, Henkelman-CMS-2006}, we place the pseudo atoms to the sites of interstitial charge maxima, and partition the charge basin by assigning the boundary according the zero flux charge density gradient between the true and pseudo nucleus. Finally, we compute the DOS and volume belonging to the interstitial sites and compare them with the totals. An ideal electride should occupy the entire volume of the crystal voids and make the primary contribution to the total DOS around $E_\textrm{F}$ (as shown in Fig. \ref{fig:scheme}). In our calculations, we define the material as an electride when the interstitial electrons occupy at least a volume ratio of 5\% at the energy range of $E_\textrm{F} \pm$ 0.05 eV.

From the available materials data in the Materials Project database \cite{MP-2013}, we made a query to obtain the first set of ICSD compounds by filtering out materials which are unlikely to be electrides (elemental solids or materials with band gap larger than 0.25 eV) or too expensive for calculation (e.g., materials containing more than 100 atoms in the primitive unit cell). This reduced the total number of candidate materials from 69640 to 17922. We then applied the computational scheme described above to search for the potential electrides from the reduced pool. The screening identified the total of 167 potential electrides, which included most of the previously discovered materials (a detailed comparison is included in Table S1). Among them, 112 were identified as electrides for the first time. These materials are made of 59 elements out of the entire periodic table. It is interesting to note that many materials can be classified as intermetallics (e.g., Ca$_3$Ag$_8$, Ca(MnAl$_2$)$_4$, Dy$_3$Co, Yb$_5$Sb$_3$, etc). This agrees with a recent HT survey \cite{burton2018high} and experimental reports \cite{intermetallic-JACS,intermetallic-anie}, suggesting that intermetallics are a largely unexplored family of materials suitable for future consideration. Another important insight that can be gained from the elements distribution is that the excess electrons are provided by elements with small electronegativity and large atomic radii, which typically appear in the alkali, alkaline earth and early transition metals in group IIIB. Among them, the alkaline earth metals (Mg: 23, Ca: 21, Sr: 17, Ba: 19) occur more frequently than others.

\begin{figure}[ht]
\centering
\includegraphics[width=0.9\linewidth]{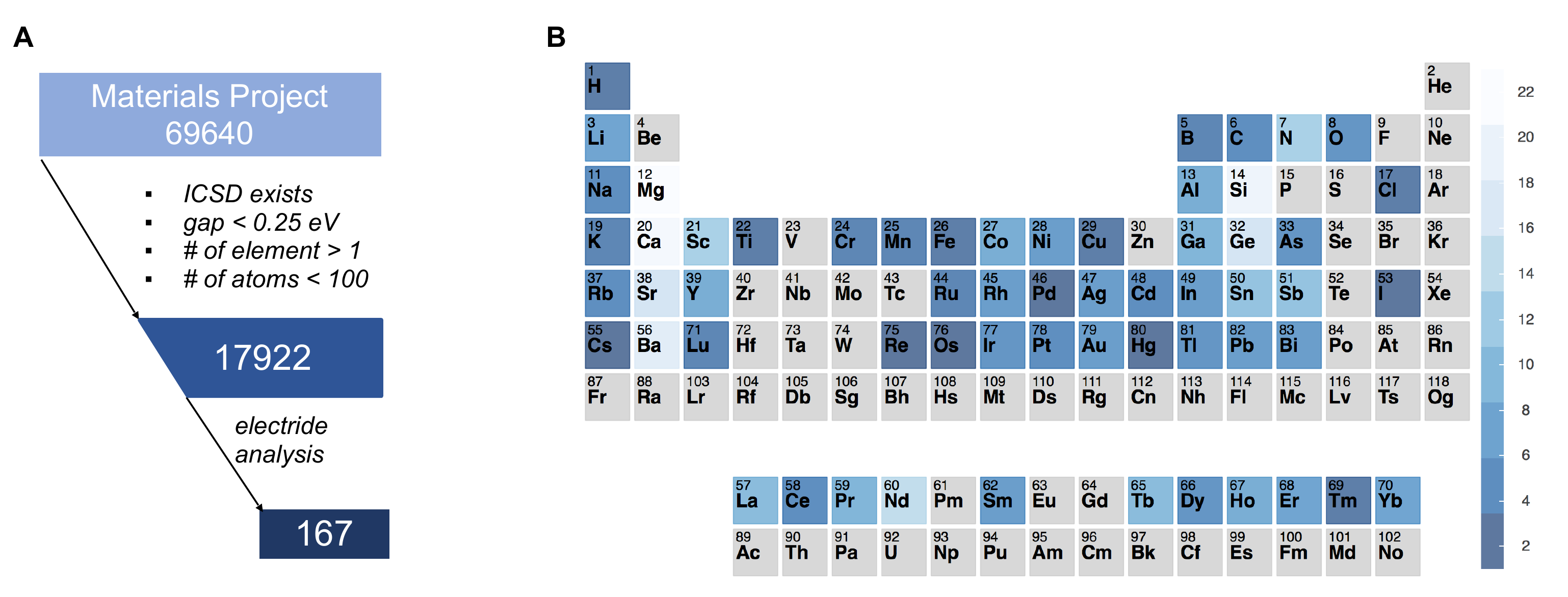}
\caption{\textbf{Computational screening of electrides from the inorganic materials database.}\\ 
(A) describes the general screening strategy based on the available materials data.\\ 
(B) shows the distribution of chemical substances among the 167 electrides identified in this work.}
\label{fig:ptable}
\end{figure}

\begin{figure}[ht]
\centering
\includegraphics[width=0.95\linewidth]{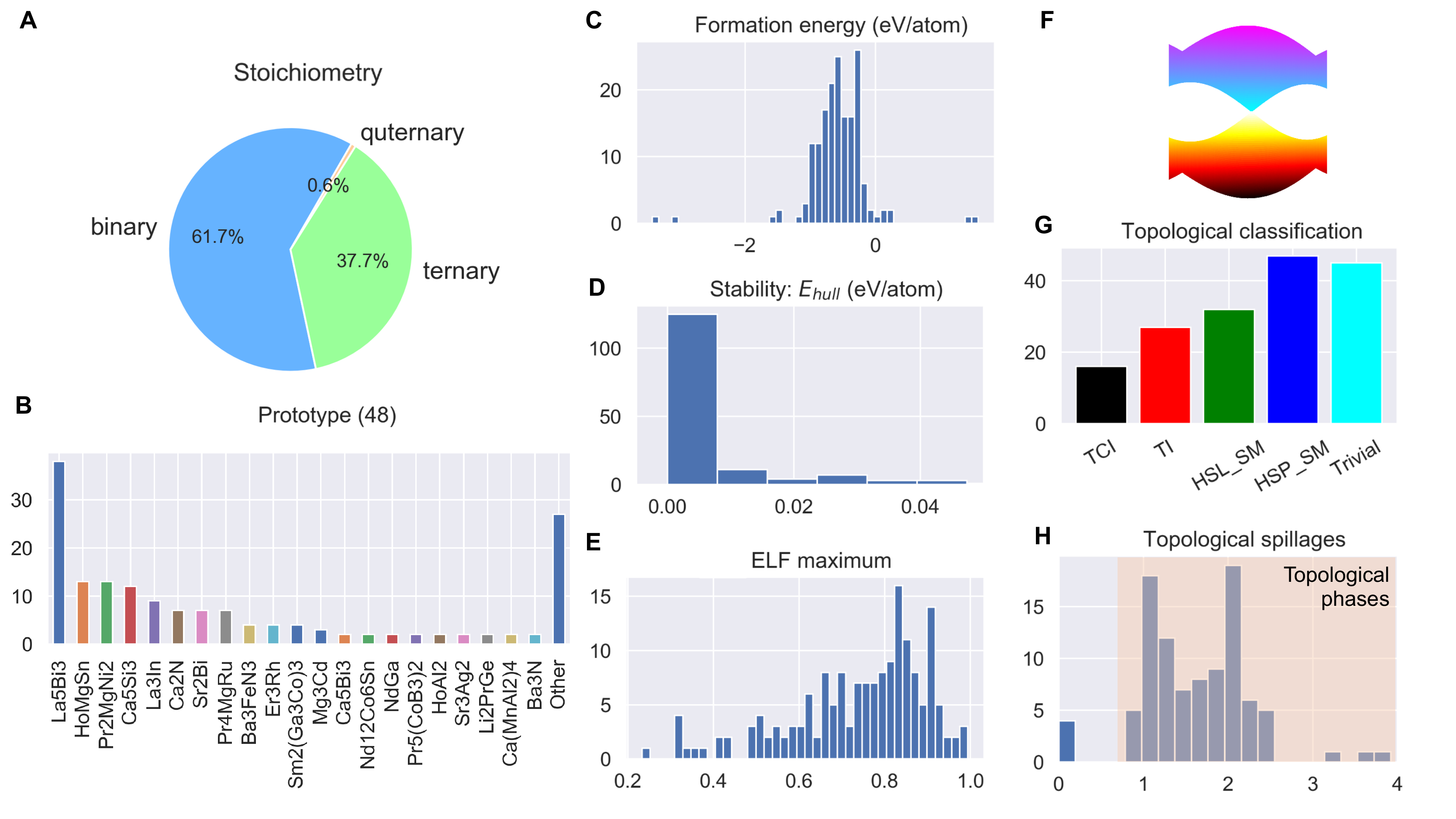}
\caption{\textbf{Statistical analysis of the selected properties of the identified 167 electride materials.}\\
(A) The distribution of stoichiometry.\\
(B) The distribution of structure prototype.\\
(C and D) The histograms of formation energy and the stability based on the energy above the convex hull. \\
(E) The histograms of ELF maximum in the crystal cavities.\\
(F) The illustration of topological band inversion.\\
(G) The classification of topological materials, including topological insulator (TI), topological crystalline insulator (TCI), high-symmetry point semi-metal(HSP\_SM), high-symmetry line semi-metal (HSL\_SM), and trivial states. \\
(H) The histogram of calculated spillage values, in which a value larger than 0.5 indicates that the material is topological.}
\label{fig:statistics}
\end{figure}

Fig. \ref{fig:statistics} shows further statistical analysis performed on the identified materials. In terms of stoichiometry, they include 103 binaries, 63 ternaries and 1 quaternaries. Since we are only interested in the materials which are recoverable at ambient conditions, the HPEs like elemental Li, Na, K, are omitted in this library. We further classify the identified electrides according to the structural similarity. We found 48 distinct prototypes, 8 of which contain more than 5 materials (Mn$_5$Si$_3$: 38, HoMgSn: 14, Pr$_2$MgNi$_2$: 13, Ca$_5$Si$_3$: 12, La$_3$In: 9, Pr$_4$MgRu: 7, Sr$_2$Bi: 7, Ca$_2$N: 7). Compared to the previous studies \cite{burton2018high,Inoshita-PRX-2014,Zhang-PRX-2017,Tada-IC-2014,walsh2013electron}, we have identified 25 new prototypes, suggesting that our descriptor is capable of discovering new classes of materials. Having identified 48 distinct prototypes, many more new electrides can be generated by chemical substitution approaches. For most materials the calculated formation energy was negative, with only few exceptions representing materials from high-pressure synthesis such as  Rb$_3$Os, Ca$_3$Pb, Na$_2$Cl, Na$_3$Cl$_2$. A better stability criterion is the normalized formation energy based on the distance to the convex hull ($E_\textrm{hull}$), which is the minimum energy of decomposition into an isochemical mixture of other phases \cite{Oganov2019}. In such convex hull construction, the stable compounds always have zero $E_\textrm{hull}$ values, while the compounds with positive $E_\textrm{hull}$ would decompose into the mixture of other neighboring compounds with zero $E_\textrm{hull}$. As shown in Fig. \ref{fig:statistics}d, the majority of compounds (116 out of 167) are thermodynamically stable against decomposition ($E_\textrm{hull}$ < 1 meV/atom). However, $E_\textrm{hull}$ itself only tells the thermodynamic preference of a compound in a closed system, the practical applications often require the room temperature stability in the presence of other environmental species (such as water and air). More extended stability analysis of these materials is a subject to future experimental investigation.

\begin{figure}[ht]
\centering
\includegraphics[width=1.0\linewidth]{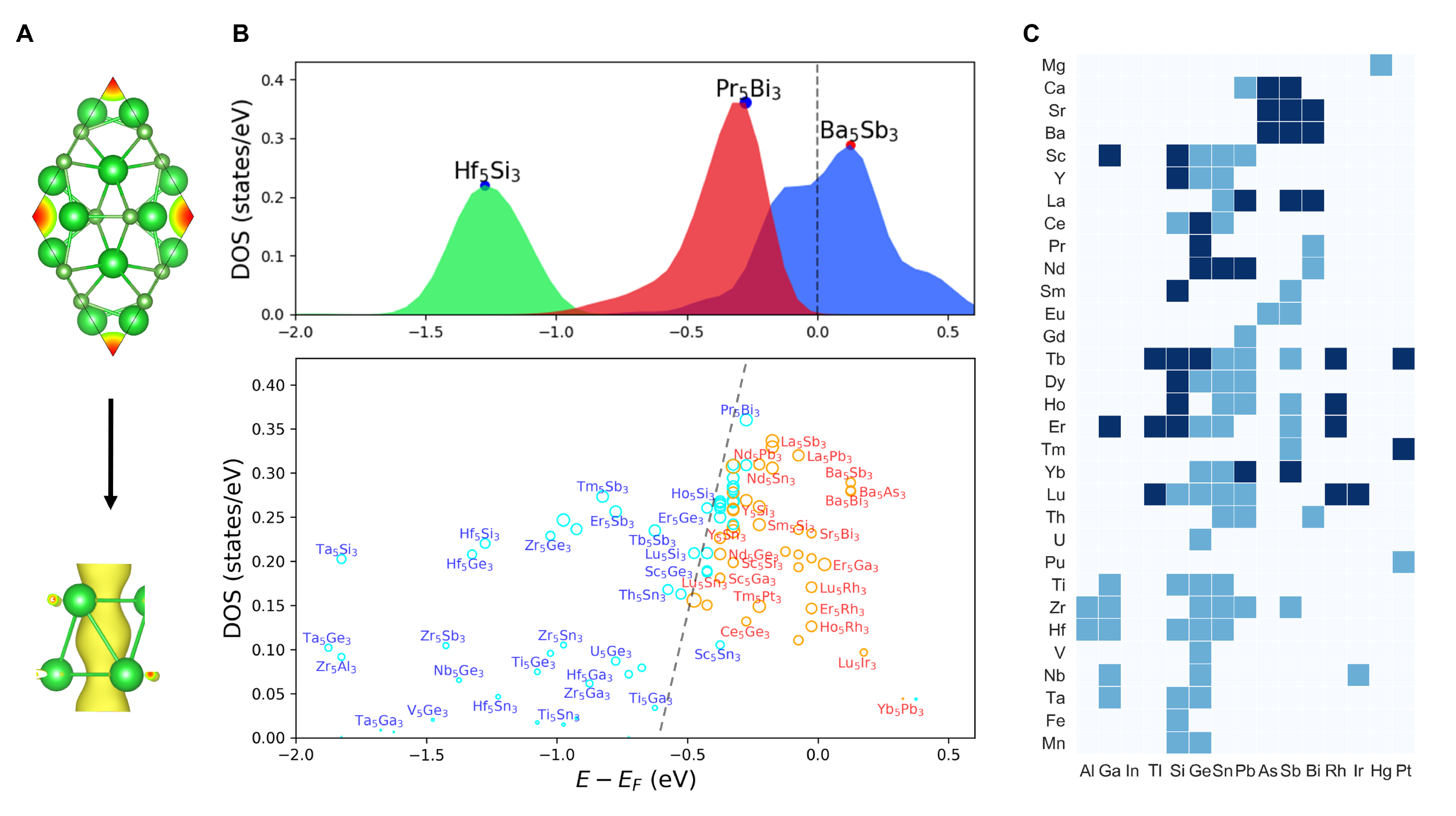}
\caption{\textbf{Statistical analysis performed for Mn$_5$Si$_3$ type compounds.}\\
(A) The interstitial electrons in the crystal Ba$_5$As$_3$ (upper panel) and its local environments (lower panel).\\ 
(B) The energy distribution of the interstitial electrons in several representative materials for three different classes, namely non-electride, potential electride and typical electride (upper panel) and the distribution of 105 Mn$_5$Si$_3$ type compounds reported in the ICSD separated by a schematic decision boundary (lower panel). In the lower panel, the cyran colored points refer to the non-electride family while the orange points refer to the electride class according to our simulation. The size of each point represent the relative Bader charge. The materials close to the decision boundary shall be considered to be potential electride class. \\
(C) The distribution of electrides in the 2D chemical space. Grey boxes indicate the absence of stable stable Mn$_5$Si$_3$ compounds for a given composition. The dark and light blue represents electrides and non-electrides, respectively.}
\label{fig:a5b3}
\end{figure}

Fig. \ref{fig:statistics}e plots the distribution of the interstitial ELF maxima (ELF$_\textrm{max}$) for the identified electride materials. Interestingly, there is a wide spread in the calculated values ranging from as low as 0.26 up to the theoretical limit of 1.0. This data suggests that using an arbitrary value of ELF maximum  \cite{Zhang-PRX-2017} to screen for electrides can miss a lot of potential candidates. On the other hand, these strong non-nuclear ELF maxima do not always correspond to the DOS around the Fermi level (as will be shown in Fig. \ref{fig:a5b3}). Although the strong ELF is not a necessary condition for an electride, the tendency of localization may be useful to predict the magnetic susceptibility. For electrides with electrons strongly localized in the dense interstitial sites, the orbital of interstitial electrons tends to behave like $d$-orbital, and thus favour magnetic ordering \cite{Pickard-PRL-2011, Miao-ACR-2014}. Indeed, some previously identified magnetic electrides have notably high ELF$_\textrm{max}$, Y$_2$C (0.87) \cite{inoshita2015ferromagnetic}, Li$_2$Ca$_3$N$_6$ (0.88) \cite{schneider2013high}, K$_4$Al$_3$(SiO$_4$)$_3$ (0.99) \cite{madsen2001electronic}. Accordingly, we expect a large number of electrides identified in this work to possess antiferro- or ferromagnetism based on the analysis shown in Fig. \ref{fig:statistics}e. 

Recent work suggested that the unique floating electrons in electrides are favorable for achieving band inversion for topological phases \cite{Hirayama-PRX-2018, Huang-NanoL-2018}. The automated classification of topological materials has been enabled recently based on the crystalline symmetries \cite{zhang2018catalogue, Vergniory2019, Tang2019}. Comparing both the topological \cite{zhang2018catalogue} and our electride materials databases, we found 122 materials belong to both catalogues (Fig. \ref{fig:statistics}f, see more details in table S1), which include all the recently reported topological electrides Ba$_3$N, Cs$_3$O, Ca$_3$Pb, Y$_2$C, Sr$_2$Bi \cite{Hirayama-PRX-2018,Huang-NanoL-2018,Ca3Pb-2018,Park-PRL-2018}. In parallel to these efforts, we adopted an alternative approach of spin-orbit spillage which has been established to identify many semimetals, topological insulators and topological crystalline insulators \cite{SOP}. Following the modified formalism used in the JARVIS-DFT \cite{choudhary2018high}, we calculated the spillage values for 96 materials based on the difference between non-spin-orbit coupling (SOC) and SOC wavefunctions for each material. A non-zero spillage value (generally larger than 0.5) indicates that the material is topological. Most of the electrides in Fig. \ref{fig:statistics}g, have high spillage values. These clearly suggests that the identified materials are topological. Trivial materials like Ca$_2$Sb (shown in table S1) have spillage values close to zero. However, there are also a few cases where symmetry indicator method \cite{zhang2018catalogue} identifies the material as trivial but the spillage such as for Ho$_4$InIr. This might be due to difference in k-mesh used in the two methodologies. The appearing strong correlation between the electrides in chemistry and the electronic band topology in condensed matter physics may be useful for designing novel functional materials with combined properties in future.

\begin{figure}[ht]
\centering
\includegraphics[width=0.8\linewidth]{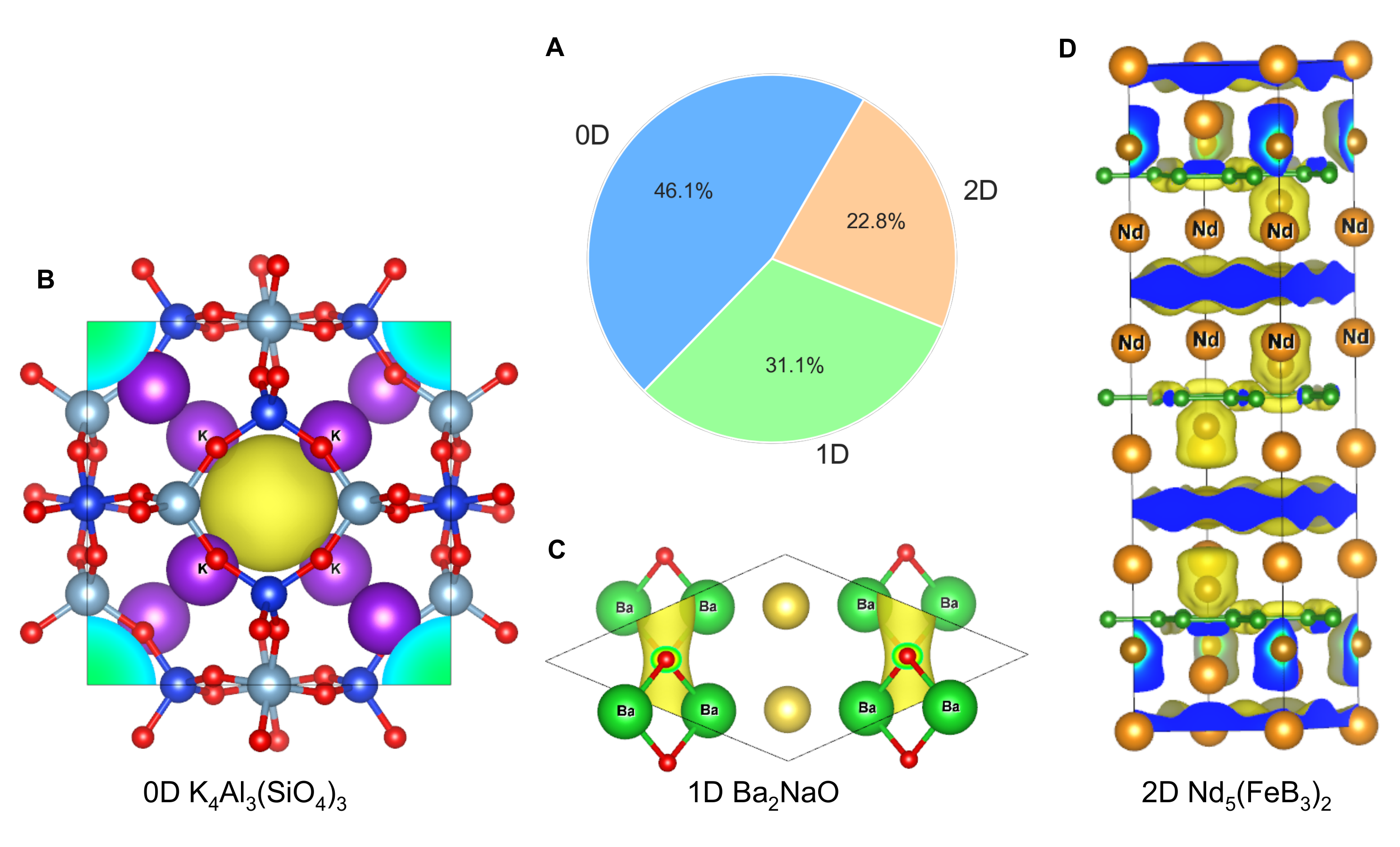}
\caption{\textbf{Example electrides categorized by the dimensionality of the interstitial electrons.}\\
(A) The distribution of dimension among all identified electride materials. \\
(B) A representative 0D electride K$_4$Al$_3$(SiO$_4$)$_3$.\\
(C) A representative 1D electride Ba$_2$NaO.\\
(D) A representative 2D electride Nd$_5$(FeB$_3$)$_2$ (2D).}
\label{fig:dimension}
\end{figure}

One of the key characteristics of electrides is the topology of the cavities confining the anionic electrons. In a zero dimensional (0D) electride (such as C12A7:2e$^-$), the anionic electrons are localized in lattice cavities and isolated from each other, while in 1D, 2D and even 3D electrides the electrons are connected along lattice channels or planes. Electrides with higher degree of connectivity are more desirable for catalytic and other applications. A variety of different geometries was found in the identified electrides, including 77 0D, 52 1D and 38 2D electrides. Remarkably, we found several compelling materials reported for the first time. For example, the pore of sodalite type K$_4$Al$_3$(SiO$_4$)$_3$ has a diameter of 1.1 nm, which is about 2-3 times larger than that of C12A7. Interestingly, we also found that Li$_{12}$Mg$_3$Si$_4$ shares a similar crystallographic packing with C12A7. It is calculated to have about 2-3 times higher excess charge densities than that of C12A7, which are thus promising to have higher mobility \cite{wang2018ternary}. Our screening procedure have identified many other 1D and 2D electides. For instance, Nd$_5$(FeB$_3$)$_2$ crystallizes in the same space group as Ca$_2$N ($R$-3$m$), representing another example of a 2D electride. We list several representative materials in Fig. \ref{fig:dimension}. The complete list can be found in the supplementary Figures S2-S49 and the supplementary table.

It is important to note that, crystal packing is not the only factor determining whether a given material is an electride. Rather, it depends on a number of different factors. We illustrate the importance of chemical composition by a comprehensive investigation of the most common electride prototype Mn$_5$Si$_3$ (A$_5$B$_3$). According to the database, we found 105 such compounds in the ICSD. Our detailed calculation of partial charge density between (-2.5$\leq E$-$E_{\textrm{F}} \leq$0 eV) suggest that 94 of them have charge maxima inside the crystal cavities with the localized electrons forming 1D arrays along the (001) direction (Fig. \ref{fig:a5b3}a). To investigate the energy distribution of these interstitial electrons, we calculated their partial DOS based on Bader partition (see supplementary materials). We found several types of PDOS profiles for these compounds as shown in Fig. \ref{fig:a5b3}b. For compounds such as Ba$_5$Sb$_3$, the PDOS peak of the interstitial electrons is very close to $E_\textrm{F}$, which is indicative of a \textit{typical electrides}. The second type of materials (such as Hf$_5$Si$_3$) has the PDOS peak at the deep energy bands away from the Fermi level. These materials are unlikely to be electrides since these excess electrons do not contribute to the DOS at $E_\textrm{F}$, and we therefore label them as \textit{non-electrides}. The third type of materials (Pr$_5$Bi$_3$) has the PDOS peak close to $E_\textrm{F}$, leading to a small portion of PDOS to be present around $E_\textrm{F}$. We categorize them as \textit{potential electrides} whose PDOS peak may be pushed to Fermi level due to external environments (e.g., pressure, temperature, chemical doping). We therefore place each compound to its PDOS peak position closest to $E_\textrm{F}$, with the size of each sphere representing the integrated PDOS value (i.e., the integrated Bader charge) between -2.0 to 0 eV. As shown in Fig. \ref{fig:a5b3}b, we can thus qualitatively draw a line to separate them by its electride properties. By analyzing their chemistry, we can clearly see that the \textit{non-electrides} group are characterized by the compounds made of late transition metals such as Ta, Zr, Hf, etc. On the other hand, the group of \textit{typical electrides} is composed of strong electron positive elements such as Ba, Sr, Y and many other lanthanides. While, the materials close to the boundary also have elements with small electronegativity values occupying both A and B sites. Note that Y$_5$Si$_3$ from this family has been recently suggested to be promising candidate material for practical application as it is water durable. But its PDOS at $E_\textrm{F}$ is relatively medium compared to other candidates. Such analysis, together with the 2D colored map (Fig. \ref{fig:a5b3}b-c) can be instructive in guiding the development of new functional electrides in future. Given that this type of materials have not yet been thoroughly explored, many more electrides are yet to be discovered in the near future.

Compared to only a few dozen of known electrides at ambient condition, we have identified 167 potential electride materials from a subset of ICSD materials. It is natural to ask why many electrides were neglected in the past. First, the traditional wisdom in electride design is to search for candidate materials from ionic solids with formal positive charges (e.g., Ca$_2$N \cite{Lee-Nature-2013}). However, we observed plenty of intermetallics in our database. This explains why a large quantity of electrides were missed in the past. Even for ionic solids, it is risky to gauge the formal charges purely from the chemical formula, as the atoms in the compound may adopt multiple valence states and adopt mixed chemical bonding \cite{wang2018ternary}. For instance, it is hard to know whether Li$_2$Ca$_3$N$_6$ is an electron rich compound without a detailed chemical bonding analysis \cite{Qu-ACSAMI-2019}. Second, some materials (e.g., Ba$_2$NaO \cite{NaBa2O-2001}, Li$_2$Ca$_3$N$_6$\cite{schneider2013high}) were characterized as electron rich compounds in the literature, while their electride properties were not examined in detail. Third, the past researches primarily focused on developing the electride materials based on chemical substitution on the known prototypes (e.g., Mn$_5$Si$_3$, Ca$_2$N). Discovering new structural prototypes was limited due to the lack of efficient computational approaches. Last, it is important to acknowledge that most of the electrides are likely to be stable at only low temperature due to the presence of highly mobile electrons. Though most of the reported materials in this work are calculated to be thermodynamically stable (zero energy above the convex hull at zero temperature) in a closed system, they may quickly react with environmental species such as water and air. Therefore, it is possible that they were neglected due to the issue of instability. We believe the large quantities of electrides from our screening will provide an invaluable guide for future study of electrides towards materials discovery with enhanced stability and physical properties.

\section*{Conclusion}
In sum, we propose a robust descriptor that can be used to identify the excess electrons in the electrides. With this methodology we search for electrides that contain interstitial electrons floating around the Fermi level by screening a large number of materials available through the ICSD. Our search yields 167 electrides including previously experimentally identified electrides. Among them, 112 new materials and 25 new structural prototypes are reported for the first time. The identified electrides show a lot of variability in their chemical composition and crystal packing, demonstrating the power of HT methods to discover novel electride. Extending this search to more materials (e.g., inorganic wide-gap semiconductors \cite{Wan2018}, insulators and organic compounds \cite{Li-Review-2016}) would be the subject of future work. In addition, we believe that the descriptor proposed in this work can be also used to identify the electride materials from more extended first-principles crystal structure search \cite{Zhang-PRX-2017, Wang-JACS-2017}. The drastically increased library of electride materials will benefit a broad range of potential applications of electrides which include catalysis, non-linear optics \cite{NLO-1, NLO-2} and electrics \cite{Kim2006}.

\section*{EXPERIMENTAL PROCEDURES}

We performed the characterization of electrides based on density functional theory (DFT) using the plane wave code VASP \cite{Vasp-PRB-1996}. Please note commercial software is identified to specify procedures. Such identification does not imply recommendation by the National Institute of Standards and Technology. The exchange-correlation functional was described by the generalized gradient approximation in the Perdew-Burke-Ernzerhof parameterization (GGA-PBE) \cite{PBE-PRL-1996}. Using the parameter sets provided by Pymatgen \cite{pymatgen-2013}, we performed geometry relaxation (MPRelaxSet) and static (MPStaticSet) calculations for each selected structure, followed by the ELF and partial charge density calculation as described in Fig. \ref{fig:scheme}. To compute the PDOS we used a reciprocal density of 250 for the first stage of screening and 500 for more accurate analysis at the second stage of characterization. The PDOS for the excess electrons are computed based on Bader grids as described in the supplementary materials (section 1). The structure prototypes were classified through the anonymous structure matcher utility in Pymatgen \cite{pymatgen-2013}. For the identification of topological materials, we searched for materials with large changes in the occupied wavefunctions due to including spin orbital coupling \cite{SOP,choudhary2018high}. Alternatively, the online database for topological materials classification (\url{http://materiae.iphy.ac.cn}) was used for further comparison.

\section*{ACKNOWLEDGMENTS}
Work at UNLV is supported by the National Nuclear Security Administration under the Stewardship Science Academic Alliances program through DOE Cooperative Agreement DE-NA0001982. Support for T.F. was provided under the auspices of the US Department of Energy by Lawrence Livermore National Laboratory under Contract DE-AC52-07NA27344. QZ acknowledge the use of computing resources from XSEDE and Center for Functional Nanomaterials under contract no. DE-AC02-98CH10086, and the inspiring discussions with Shyueping Ong, Junjie Wang, Wei Ren and Heng Gao.

\bibliography{sample}

\begin{thebibliography}{10}
\urlstyle{rm}
\expandafter\ifx\csname url\endcsname\relax
  \def\url#1{\texttt{#1}}\fi
\expandafter\ifx\csname urlprefix\endcsname\relax\def\urlprefix{URL }\fi
\expandafter\ifx\csname doiprefix\endcsname\relax\def\doiprefix{DOI: }\fi
\providecommand{\bibinfo}[2]{#2}
\providecommand{\eprint}[2][]{\url{#2}}

\bibitem{Dye-Science-1990}
\bibinfo{author}{Dye, J.~L.}
\newblock \bibinfo{journal}{\bibinfo{title}{Electrides: Ionic salts with
  electrons as the anions}}.
\newblock {\emph{\JournalTitle{Science}}} \textbf{\bibinfo{volume}{247}},
  \bibinfo{pages}{663--668}, \doiprefix\url{10.1126/science.247.4943.663}
  (\bibinfo{year}{1990}).

\bibitem{dye2009electrides}
\bibinfo{author}{Dye, J.~L.}
\newblock \bibinfo{journal}{\bibinfo{title}{Electrides: early examples of
  quantum confinement}}.
\newblock {\emph{\JournalTitle{Acc. Chem. Res.}}}
  \textbf{\bibinfo{volume}{42}}, \bibinfo{pages}{1564--1572},
  \doiprefix\url{10.1021/ar9000857} (\bibinfo{year}{2009}).

\bibitem{Ellaboudy-1983-JACS}
\bibinfo{author}{Ellaboudy, A.}, \bibinfo{author}{Dye, J.~L.} \&
  \bibinfo{author}{Smith, P.~B.}
\newblock \bibinfo{journal}{\bibinfo{title}{Cesium 18-crown-6 compounds. a
  crystalline ceside and a crystalline electride}}.
\newblock {\emph{\JournalTitle{J. Am. Chem. Soc.}}}
  \textbf{\bibinfo{volume}{105}}, \bibinfo{pages}{6490--6491},
  \doiprefix\url{10.1021/ja00359a022} (\bibinfo{year}{1983}).

\bibitem{Matsuishi-Science-2003}
\bibinfo{author}{Matsuishi, S.} \emph{et~al.}
\newblock \bibinfo{journal}{\bibinfo{title}{High-density electron anions in a
  nanoporous single crystal: {[Ca$_{24}$Al$_{28}$O$_{64}$]$^{4+}$(4$e^-$)}}}.
\newblock {\emph{\JournalTitle{Science}}} \textbf{\bibinfo{volume}{301}},
  \bibinfo{pages}{626--629}, \doiprefix\url{10.1126/science.1083842}
  (\bibinfo{year}{2003}).

\bibitem{Kitano-NChem-2012}
\bibinfo{author}{Kitano, M.} \emph{et~al.}
\newblock \bibinfo{journal}{\bibinfo{title}{Ammonia synthesis using a stable
  electride as an electron donor and reversible hydrogen store}}.
\newblock {\emph{\JournalTitle{Nat. Chem.}}} \textbf{\bibinfo{volume}{4}},
  \bibinfo{pages}{934--940}, \doiprefix\url{10.1038/nchem.1476}
  (\bibinfo{year}{2012}).

\bibitem{Kuganathan-JACS-2014}
\bibinfo{author}{Kuganathan, N.}, \bibinfo{author}{Hosono, H.},
  \bibinfo{author}{Shluger, A.~L.} \& \bibinfo{author}{Sushko, P.~V.}
\newblock \bibinfo{journal}{\bibinfo{title}{Enhanced {N$_2$} dissociation on
  {Ru}-loaded inorganic electride}}.
\newblock {\emph{\JournalTitle{J. Am. Chem. Soc.}}}
  \textbf{\bibinfo{volume}{136}}, \bibinfo{pages}{2216--2219},
  \doiprefix\url{10.1021/ja410925g} (\bibinfo{year}{2014}).

\bibitem{Hayashi-JACS-2014}
\bibinfo{author}{Hayashi, F.} \emph{et~al.}
\newblock \bibinfo{journal}{\bibinfo{title}{Nh2--dianion entrapped in a
  nanoporous {12CaO $\cdot$ 7Al$_2$O$_3$} crystal by ammonothermal treatment:
  Reaction pathways, dynamics, and chemical stability}}.
\newblock {\emph{\JournalTitle{J. Am. Chem. Soc.}}}
  \textbf{\bibinfo{volume}{136}}, \bibinfo{pages}{11698--11706},
  \doiprefix\url{10.1021/ja504185m} (\bibinfo{year}{2014}).

\bibitem{Hosono-PNAS-2017}
\bibinfo{author}{Hosono, H.}, \bibinfo{author}{Kim, J.}, \bibinfo{author}{Toda,
  Y.}, \bibinfo{author}{Kamiya, T.} \& \bibinfo{author}{Watanabe, S.}
\newblock \bibinfo{journal}{\bibinfo{title}{Transparent amorphous oxide
  semiconductors for organic electronics: Application to inverted oleds}}.
\newblock {\emph{\JournalTitle{Proc. Natl. Acad. Sci.}}}
  \textbf{\bibinfo{volume}{114}}, \bibinfo{pages}{233--238},
  \doiprefix\url{10.1073/pnas.1617186114} (\bibinfo{year}{2017}).

\bibitem{li2004theoretical}
\bibinfo{author}{Li, H.} \& \bibinfo{author}{Mahanti, S.}
\newblock \bibinfo{journal}{\bibinfo{title}{Theoretical study of encapsulated
  alkali metal atoms in nanoporous channels of {ITQ}-4 zeolite: One-dimensional
  metals and inorganic electrides}}.
\newblock {\emph{\JournalTitle{Phys. Rev. Lett.}}}
  \textbf{\bibinfo{volume}{93}}, \bibinfo{pages}{216406},
  \doiprefix\url{10.1103/PhysRevLett.93.216406} (\bibinfo{year}{2004}).

\bibitem{dye2005alkali}
\bibinfo{author}{Dye, J.~L.} \emph{et~al.}
\newblock \bibinfo{journal}{\bibinfo{title}{Alkali metals plus silica gel:
  powerful reducing agents and convenient hydrogen sources}}.
\newblock {\emph{\JournalTitle{J. Am. Chem. Soc.}}}
  \textbf{\bibinfo{volume}{127}}, \bibinfo{pages}{9338--9339},
  \doiprefix\url{10.1021/ja051786\%2B} (\bibinfo{year}{2005}).

\bibitem{Lee-Nature-2013}
\bibinfo{author}{Lee, K.}, \bibinfo{author}{Kim, S.~W.}, \bibinfo{author}{Toda,
  Y.}, \bibinfo{author}{Matsuishi, S.} \& \bibinfo{author}{Hosono, H.}
\newblock \bibinfo{journal}{\bibinfo{title}{Dicalcium nitride as a
  two-dimensional electride with an anionic electron layer}}.
\newblock {\emph{\JournalTitle{Nature}}} \textbf{\bibinfo{volume}{494}},
  \bibinfo{pages}{336--340}, \doiprefix\url{10.1038/nature11812}
  (\bibinfo{year}{2013}).

\bibitem{Zhang-JPCL-2015}
\bibinfo{author}{Zhang, Y.}, \bibinfo{author}{Xiao, Z.},
  \bibinfo{author}{Kamiya, T.} \& \bibinfo{author}{Hosono, H.}
\newblock \bibinfo{journal}{\bibinfo{title}{Electron confinement in channel
  spaces for one-dimensional electride}}.
\newblock {\emph{\JournalTitle{J. Phys. Chem. Lett.}}}
  \textbf{\bibinfo{volume}{6}}, \bibinfo{pages}{4966--4971},
  \doiprefix\url{10.1021/acs.jpclett.5b02283} (\bibinfo{year}{2015}).

\bibitem{Zhang-QM-2017}
\bibinfo{author}{Zhang, Y.} \emph{et~al.}
\newblock \bibinfo{journal}{\bibinfo{title}{Electride and superconductivity
  behaviors in {Mn$_5$Si$_3$}-type intermetallics}}.
\newblock {\emph{\JournalTitle{npj Quantum Materials}}}
  \textbf{\bibinfo{volume}{2}}, \bibinfo{pages}{45},
  \doiprefix\url{10.1038/s41535-017-0053-4} (\bibinfo{year}{2017}).

\bibitem{Wang-JACS-2017}
\bibinfo{author}{Wang, J.} \emph{et~al.}
\newblock \bibinfo{journal}{\bibinfo{title}{Exploration of stable strontium
  phosphide-based electrides: Theoretical structure prediction and experimental
  validation}}.
\newblock {\emph{\JournalTitle{J. Am. Chem. Soc.}}}
  \textbf{\bibinfo{volume}{139}}, \bibinfo{pages}{15668--15680},
  \doiprefix\url{10.1021/jacs.7b06279} (\bibinfo{year}{2017}).

\bibitem{Lu-JACS-2016}
\bibinfo{author}{Lu, Y.} \emph{et~al.}
\newblock \bibinfo{journal}{\bibinfo{title}{Water durable electride
  {Y$_5$Si$_3$}: Electronic structure and catalytic activity for ammonia
  synthesis}}.
\newblock {\emph{\JournalTitle{J. Am. Chem. Soc.}}}
  \textbf{\bibinfo{volume}{138}}, \bibinfo{pages}{3970--3973},
  \doiprefix\url{10.1021/jacs.6b00124} (\bibinfo{year}{2016}).

\bibitem{Inoshita-PRX-2014}
\bibinfo{author}{Inoshita, T.}, \bibinfo{author}{Jeong, S.},
  \bibinfo{author}{Hamada, N.} \& \bibinfo{author}{Hosono, H.}
\newblock \bibinfo{journal}{\bibinfo{title}{Exploration for two-dimensional
  electrides via database screening and ab initio calculation}}.
\newblock {\emph{\JournalTitle{Phys. Rev. X}}} \textbf{\bibinfo{volume}{4}},
  \bibinfo{pages}{031023}, \doiprefix\url{10.1103/PhysRevX.4.031023}
  (\bibinfo{year}{2014}).

\bibitem{Tada-IC-2014}
\bibinfo{author}{Tada, T.}, \bibinfo{author}{Takemoto, S.},
  \bibinfo{author}{Matsuishi, S.} \& \bibinfo{author}{Hosono, H.}
\newblock \bibinfo{journal}{\bibinfo{title}{High-throughput ab initio screening
  for two-dimensional electride materials}}.
\newblock {\emph{\JournalTitle{Inorg. Chem.}}} \textbf{\bibinfo{volume}{53}},
  \bibinfo{pages}{10347--10358}, \doiprefix\url{10.1021/ic501362b}
  (\bibinfo{year}{2014}).

\bibitem{Zhang-PRX-2017}
\bibinfo{author}{Zhang, Y.}, \bibinfo{author}{Wang, H.}, \bibinfo{author}{Wang,
  Y.}, \bibinfo{author}{Zhang, L.} \& \bibinfo{author}{Ma, Y.}
\newblock \bibinfo{journal}{\bibinfo{title}{Computer-assisted inverse design of
  inorganic electrides}}.
\newblock {\emph{\JournalTitle{Phys. Rev. X}}} \textbf{\bibinfo{volume}{7}},
  \bibinfo{pages}{011017}, \doiprefix\url{10.1103/PhysRevX.7.011017}
  (\bibinfo{year}{2017}).

\bibitem{burton2018high}
\bibinfo{author}{Burton, L.~A.}, \bibinfo{author}{Ricci, F.},
  \bibinfo{author}{Chen, W.}, \bibinfo{author}{Rignanese, G.-M.} \&
  \bibinfo{author}{Hautier, G.}
\newblock \bibinfo{journal}{\bibinfo{title}{High-throughput identification of
  electrides from all known inorganic materials}}.
\newblock {\emph{\JournalTitle{Chem. Mater.}}} \textbf{\bibinfo{volume}{30}},
  \bibinfo{pages}{7521--7526}, \doiprefix\url{10.1021/acs.chemmater.8b02526}
  (\bibinfo{year}{2018}).

\bibitem{Pickard-PRL-2011}
\bibinfo{author}{Pickard, C.~J.} \& \bibinfo{author}{Needs, R.~J.}
\newblock \bibinfo{journal}{\bibinfo{title}{Predicted pressure-induced $s$-band
  ferromagnetism in alkali metals}}.
\newblock {\emph{\JournalTitle{Phys. Rev. Lett.}}}
  \textbf{\bibinfo{volume}{107}}, \bibinfo{pages}{087201},
  \doiprefix\url{10.1103/PhysRevLett.107.087201} (\bibinfo{year}{2011}).

\bibitem{inoshita2015ferromagnetic}
\bibinfo{author}{Inoshita, T.}, \bibinfo{author}{Hamada, N.} \&
  \bibinfo{author}{Hosono, H.}
\newblock \bibinfo{journal}{\bibinfo{title}{Ferromagnetic instability of
  interlayer floating electrons in the quasi-two-dimensional electride
  {Y$_2$C}}}.
\newblock {\emph{\JournalTitle{Phys. Rev. B}}} \textbf{\bibinfo{volume}{92}},
  \bibinfo{pages}{201109}, \doiprefix\url{10.1103/PhysRevB.92.201109}
  (\bibinfo{year}{2015}).

\bibitem{Huang-NanoL-2018}
\bibinfo{author}{Huang, H.}, \bibinfo{author}{Jin, K.-H.},
  \bibinfo{author}{Zhang, S.} \& \bibinfo{author}{Liu, F.}
\newblock \bibinfo{journal}{\bibinfo{title}{Topological electride {Y$_2$C}}}.
\newblock {\emph{\JournalTitle{Nano Lett.}}} \textbf{\bibinfo{volume}{18}},
  \bibinfo{pages}{1972--1977}, \doiprefix\url{10.1021/acs.nanolett.7b05386}
  (\bibinfo{year}{2018}).

\bibitem{Hirayama-PRX-2018}
\bibinfo{author}{Hirayama, M.}, \bibinfo{author}{Matsuishi, S.},
  \bibinfo{author}{Hosono, H.} \& \bibinfo{author}{Murakami, S.}
\newblock \bibinfo{journal}{\bibinfo{title}{Electrides as a new platform of
  topological materials}}.
\newblock {\emph{\JournalTitle{Phys. Rev. X}}} \textbf{\bibinfo{volume}{8}},
  \bibinfo{pages}{031067}, \doiprefix\url{10.1103/PhysRevX.8.031067}
  (\bibinfo{year}{2018}).

\bibitem{Ca3Pb-2018}
\bibinfo{author}{Zhang, X.}, \bibinfo{author}{Guo, R.}, \bibinfo{author}{Jin,
  L.}, \bibinfo{author}{Dai, X.} \& \bibinfo{author}{Liu, G.}
\newblock \bibinfo{journal}{\bibinfo{title}{Intermetallic {Ca$_3$Pb}: a
  topological zero-dimensional electride material}}.
\newblock {\emph{\JournalTitle{J. Mater. Chem. C}}}
  \textbf{\bibinfo{volume}{6}}, \bibinfo{pages}{575--581},
  \doiprefix\url{10.1039/C7TC04989G} (\bibinfo{year}{2018}).

\bibitem{Park-PRL-2018}
\bibinfo{author}{Park, C.}, \bibinfo{author}{Kim, S.~W.} \&
  \bibinfo{author}{Yoon, M.}
\newblock \bibinfo{journal}{\bibinfo{title}{First-principles prediction of new
  electrides with nontrivial band topology based on one-dimensional building
  blocks}}.
\newblock {\emph{\JournalTitle{Phys. Rev. Lett.}}}
  \textbf{\bibinfo{volume}{120}}, \bibinfo{pages}{026401},
  \doiprefix\url{10.1103/PhysRevLett.120.026401} (\bibinfo{year}{2018}).

\bibitem{Zhu-PRM-2019}
\bibinfo{author}{Zhu, S.-C.} \emph{et~al.}
\newblock \bibinfo{journal}{\bibinfo{title}{Computational design of flexible
  electrides with nontrivial band topology}}.
\newblock {\emph{\JournalTitle{Phys. Rev. Materials}}}
  \textbf{\bibinfo{volume}{3}}, \bibinfo{pages}{024205},
  \doiprefix\url{10.1103/PhysRevMaterials.3.024205} (\bibinfo{year}{2019}).

\bibitem{Pickard-PRL-2009}
\bibinfo{author}{Pickard, C.~J.} \& \bibinfo{author}{Needs, R.}
\newblock \bibinfo{journal}{\bibinfo{title}{Dense low-coordination phases of
  lithium}}.
\newblock {\emph{\JournalTitle{Phys. Rev. Lett.}}}
  \textbf{\bibinfo{volume}{102}}, \bibinfo{pages}{146401},
  \doiprefix\url{10.1103/PhysRevLett.102.146401} (\bibinfo{year}{2009}).

\bibitem{Matsuoka-Nature-2009}
\bibinfo{author}{Matsuoka, T.} \& \bibinfo{author}{Shimizu, K.}
\newblock \bibinfo{journal}{\bibinfo{title}{Direct observation of a
  pressure-induced metal-to-semiconductor transition in lithium}}.
\newblock {\emph{\JournalTitle{Nature}}} \textbf{\bibinfo{volume}{458}},
  \bibinfo{pages}{186}, \doiprefix\url{10.1038/nature07827}
  (\bibinfo{year}{2009}).

\bibitem{Ma-Nature-2009}
\bibinfo{author}{Ma, Y.} \emph{et~al.}
\newblock \bibinfo{journal}{\bibinfo{title}{Transparent dense sodium}}.
\newblock {\emph{\JournalTitle{Nature}}} \textbf{\bibinfo{volume}{458}},
  \bibinfo{pages}{182}, \doiprefix\url{10.1038/nature07786}
  (\bibinfo{year}{2009}).

\bibitem{Dong-NChem-2017}
\bibinfo{author}{Dong, X.} \emph{et~al.}
\newblock \bibinfo{journal}{\bibinfo{title}{A stable compound of helium and
  sodium at high pressure}}.
\newblock {\emph{\JournalTitle{Nat. Chem.}}} \textbf{\bibinfo{volume}{9}},
  \bibinfo{pages}{440}, \doiprefix\url{10.1038/nchem.2716}
  (\bibinfo{year}{2017}).

\bibitem{Zhu-PCCP-2013}
\bibinfo{author}{Zhu, Q.}, \bibinfo{author}{Oganov, A.~R.} \&
  \bibinfo{author}{Lyakhov, A.~O.}
\newblock \bibinfo{journal}{\bibinfo{title}{Novel stable compounds in the
  {Mg-O} system under high pressure}}.
\newblock {\emph{\JournalTitle{Phys. Chem. Chem. Phys.}}}
  \textbf{\bibinfo{volume}{15}}, \bibinfo{pages}{7696--7700},
  \doiprefix\url{10.1039/C3CP50678A} (\bibinfo{year}{2013}).

\bibitem{Lv-PRL-2011}
\bibinfo{author}{Lv, J.}, \bibinfo{author}{Wang, Y.}, \bibinfo{author}{Zhu, L.}
  \& \bibinfo{author}{Ma, Y.}
\newblock \bibinfo{journal}{\bibinfo{title}{Predicted novel high-pressure
  phases of lithium}}.
\newblock {\emph{\JournalTitle{Phys. Rev. Lett.}}}
  \textbf{\bibinfo{volume}{106}}, \bibinfo{pages}{015503},
  \doiprefix\url{10.1103/PhysRevLett.106.015503} (\bibinfo{year}{2011}).

\bibitem{Shimizu-Nature-2002}
\bibinfo{author}{Shimizu, K.}, \bibinfo{author}{Ishikawa, H.},
  \bibinfo{author}{Takao, D.}, \bibinfo{author}{Yagi, T.} \&
  \bibinfo{author}{Amaya, K.}
\newblock \bibinfo{journal}{\bibinfo{title}{Superconductivity in compressed
  lithium at 20 k}}.
\newblock {\emph{\JournalTitle{Nature}}} \textbf{\bibinfo{volume}{419}},
  \bibinfo{pages}{597}, \doiprefix\url{10.1038/nature01098}
  (\bibinfo{year}{2002}).

\bibitem{ICSD}
\bibinfo{author}{Bergerhoff, G.}, \bibinfo{author}{Brown, I.},
  \bibinfo{author}{Allen, F.} \emph{et~al.}
\newblock \bibinfo{journal}{\bibinfo{title}{Crystallographic databases}}.
\newblock {\emph{\JournalTitle{International Union of Crystallography,
  Chester}}} \textbf{\bibinfo{volume}{360}}, \bibinfo{pages}{77--95}
  (\bibinfo{year}{1987}).

\bibitem{Curtarolo-2013}
\bibinfo{author}{Curtarolo, S.} \emph{et~al.}
\newblock \bibinfo{journal}{\bibinfo{title}{The high-throughput highway to
  computational materials design}}.
\newblock {\emph{\JournalTitle{Nat. Mater.}}} \textbf{\bibinfo{volume}{12}},
  \bibinfo{pages}{191}, \doiprefix\url{10.1038/nmat3568}
  (\bibinfo{year}{2013}).

\bibitem{MP-2013}
\bibinfo{author}{Jain, A.} \emph{et~al.}
\newblock \bibinfo{journal}{\bibinfo{title}{Commentary: The materials project:
  A materials genome approach to accelerating materials innovation}}.
\newblock {\emph{\JournalTitle{Apl Materials}}} \textbf{\bibinfo{volume}{1}},
  \bibinfo{pages}{011002}, \doiprefix\url{10.1063/1.4812323}
  (\bibinfo{year}{2013}).

\bibitem{Li-Review-2016}
\bibinfo{author}{Zhao, S.}, \bibinfo{author}{Kan, E.} \& \bibinfo{author}{Li,
  Z.}
\newblock \bibinfo{journal}{\bibinfo{title}{Electride: from computational
  characterization to theoretical design}}.
\newblock {\emph{\JournalTitle{Wiley Interdisciplinary Reviews: Computational
  Molecular Science}}} \textbf{\bibinfo{volume}{6}}, \bibinfo{pages}{430--440},
  \doiprefix\url{10.1002/wcms.1258} (\bibinfo{year}{2016}).

\bibitem{dale-2018}
\bibinfo{author}{Dale, S.~G.} \& \bibinfo{author}{Johnson, E.~R.}
\newblock \bibinfo{journal}{\bibinfo{title}{Theoretical descriptors of
  electrides}}.
\newblock {\emph{\JournalTitle{J. Phys. Chem. A}}}
  \textbf{\bibinfo{volume}{122}}, \bibinfo{pages}{9371--9391},
  \doiprefix\url{10.1021/acs.jpca.8b08548} (\bibinfo{year}{2018}).

\bibitem{ELF}
\bibinfo{author}{Becke, A.~D.} \& \bibinfo{author}{Edgecombe, K.~E.}
\newblock \bibinfo{journal}{\bibinfo{title}{A simple measure of electron
  localization in atomic and molecular systems}}.
\newblock {\emph{\JournalTitle{J. Chem. Phys.}}} \textbf{\bibinfo{volume}{92}},
  \bibinfo{pages}{5397--5403}, \doiprefix\url{10.1063/1.458517}
  (\bibinfo{year}{1990}).

\bibitem{Bader-1990}
\bibinfo{author}{Bader, R. F.~W.}
\newblock \emph{\bibinfo{title}{Atoms in Molecules - A Quantum Theory}}
  (\bibinfo{publisher}{Oxford University Press}, \bibinfo{year}{1990}).

\bibitem{Henkelman-CMS-2006}
\bibinfo{author}{Henkelman, G.}, \bibinfo{author}{Arnaldsson, A.} \&
  \bibinfo{author}{Jonsson, H.}
\newblock \bibinfo{journal}{\bibinfo{title}{A fast and robust algorithm for
  bader decomposition of charge density}}.
\newblock {\emph{\JournalTitle{Comput. Mater. Sci.}}}
  \textbf{\bibinfo{volume}{36}}, \bibinfo{pages}{354 -- 360},
  \doiprefix\url{10.1016/j.commatsci.2005.04.010} (\bibinfo{year}{2006}).

\bibitem{intermetallic-JACS}
\bibinfo{author}{Mizoguchi, H.} \emph{et~al.}
\newblock \bibinfo{journal}{\bibinfo{title}{Zeolitic intermetallics: Lnnisi (ln
  = la–nd)}}.
\newblock {\emph{\JournalTitle{J. Am. Chem. Soc.}}}
  \textbf{\bibinfo{volume}{141}}, \bibinfo{pages}{3376--3379},
  \doiprefix\url{10.1021/jacs.8b12784} (\bibinfo{year}{2019}).

\bibitem{intermetallic-anie}
\bibinfo{author}{Wu, J.} \emph{et~al.}
\newblock \bibinfo{journal}{\bibinfo{title}{Intermetallic electride catalyst as
  a platform for ammonia synthesis}}.
\newblock {\emph{\JournalTitle{Angew. Chem. Int. Ed.}}}
  \textbf{\bibinfo{volume}{58}}, \bibinfo{pages}{825--829},
  \doiprefix\url{10.1002/anie.201812131} (\bibinfo{year}{2019}).

\bibitem{walsh2013electron}
\bibinfo{author}{Walsh, A.} \& \bibinfo{author}{Scanlon, D.~O.}
\newblock \bibinfo{journal}{\bibinfo{title}{Electron excess in alkaline earth
  sub-nitrides: 2d electron gas or 3d electride?}}
\newblock {\emph{\JournalTitle{J. Mater. Chem. C}}}
  \textbf{\bibinfo{volume}{1}}, \bibinfo{pages}{3525--3528},
  \doiprefix\url{10.1039/C3TC30690A} (\bibinfo{year}{2013}).

\bibitem{Oganov2019}
\bibinfo{author}{Oganov, A.~R.}, \bibinfo{author}{Pickard, C.~J.},
  \bibinfo{author}{Zhu, Q.} \& \bibinfo{author}{Needs, R.~J.}
\newblock \bibinfo{journal}{\bibinfo{title}{{Structure prediction drives
  materials discovery}}}.
\newblock {\emph{\JournalTitle{Nature Reviews Materials}}}
  \doiprefix\url{10.1038/s41578-019-0101-8} (\bibinfo{year}{2019}).

\bibitem{Miao-ACR-2014}
\bibinfo{author}{Miao, M.-S.} \& \bibinfo{author}{Hoffmann, R.}
\newblock \bibinfo{journal}{\bibinfo{title}{High pressure electrides: A
  predictive chemical and physical theory}}.
\newblock {\emph{\JournalTitle{Acc. Chem. Res.}}}
  \textbf{\bibinfo{volume}{47}}, \bibinfo{pages}{1311--1317},
  \doiprefix\url{10.1021/ar4002922} (\bibinfo{year}{2014}).

\bibitem{schneider2013high}
\bibinfo{author}{Schneider, S.~B.} \emph{et~al.}
\newblock \bibinfo{journal}{\bibinfo{title}{High-pressure synthesis and
  characterization of {Li$_2$Ca$_3$[N$_2$]$_3$}: An uncommon metallic diazenide
  with {[N$_2$]$^{2-}$ Ions}}}.
\newblock {\emph{\JournalTitle{J. Am. Chem. Soc}}}
  \textbf{\bibinfo{volume}{135}}, \bibinfo{pages}{16668--16679},
  \doiprefix\url{10.1021/ja408816t} (\bibinfo{year}{2013}).

\bibitem{madsen2001electronic}
\bibinfo{author}{Madsen, G.~K.}, \bibinfo{author}{Iversen, B.~B.},
  \bibinfo{author}{Blaha, P.} \& \bibinfo{author}{Schwarz, K.}
\newblock \bibinfo{journal}{\bibinfo{title}{Electronic structure of the sodium
  and potassium electrosodalites {(Na/K)$_8$(AlSiO$_4$)$_6$}}}.
\newblock {\emph{\JournalTitle{Phys. Rev. B}}} \textbf{\bibinfo{volume}{64}},
  \bibinfo{pages}{195102}, \doiprefix\url{10.1103/PhysRevB.64.195102}
  (\bibinfo{year}{2001}).

\bibitem{zhang2018catalogue}
\bibinfo{author}{Zhang, T.} \emph{et~al.}
\newblock \bibinfo{journal}{\bibinfo{title}{Catalogue of topological electronic
  materials}}.
\newblock {\emph{\JournalTitle{Nature}}} \textbf{\bibinfo{volume}{566}},
  \bibinfo{pages}{475--479}, \doiprefix\url{10.1038/s41586-019-0944-6}
  (\bibinfo{year}{2019}).

\bibitem{Vergniory2019}
\bibinfo{author}{Vergniory, M.~G.} \emph{et~al.}
\newblock \bibinfo{journal}{\bibinfo{title}{{A complete catalogue of
  high-quality topological materials}}}.
\newblock {\emph{\JournalTitle{Nature}}} \textbf{\bibinfo{volume}{566}},
  \bibinfo{pages}{480--485}, \doiprefix\url{10.1038/s41586-019-0954-4}
  (\bibinfo{year}{2019}).

\bibitem{Tang2019}
\bibinfo{author}{Tang, F.}, \bibinfo{author}{Po, H.~C.},
  \bibinfo{author}{Vishwanath, A.} \& \bibinfo{author}{Wan, X.}
\newblock \bibinfo{journal}{\bibinfo{title}{{Comprehensive search for
  topological materials using symmetry indicators}}}.
\newblock {\emph{\JournalTitle{Nature}}} \textbf{\bibinfo{volume}{566}},
  \bibinfo{pages}{486--489}, \doiprefix\url{10.1038/s41586-019-0937-5}
  (\bibinfo{year}{2019}).

\bibitem{SOP}
\bibinfo{author}{Liu, J.} \& \bibinfo{author}{Vanderbilt, D.}
\newblock \bibinfo{journal}{\bibinfo{title}{Spin-orbit spillage as a measure of
  band inversion in insulators}}.
\newblock {\emph{\JournalTitle{Phys. Rev. B}}} \textbf{\bibinfo{volume}{90}},
  \bibinfo{pages}{125133}, \doiprefix\url{10.1103/PhysRevB.90.125133}
  (\bibinfo{year}{2014}).

\bibitem{choudhary2018high}
\bibinfo{author}{Choudhary, K.}, \bibinfo{author}{Garrity, K.~F.} \&
  \bibinfo{author}{Tavazza, F.}
\newblock \bibinfo{journal}{\bibinfo{title}{High-throughput discovery of
  topological materials using spin-orbit spillage}}.
\newblock {\emph{\JournalTitle{arXiv preprint arXiv:1810.10640}}}
  (\bibinfo{year}{2018}).

\bibitem{wang2018ternary}
\bibinfo{author}{Wang, J.}, \bibinfo{author}{Zhu, Q.}, \bibinfo{author}{Wang,
  Z.} \& \bibinfo{author}{Hosono, H.}
\newblock \bibinfo{journal}{\bibinfo{title}{Ternary inorganic electrides with
  mixed bonding}}.
\newblock {\emph{\JournalTitle{Phys. Rev. B}}} \textbf{\bibinfo{volume}{99}},
  \bibinfo{pages}{064104}, \doiprefix\url{10.1103/PhysRevB.99.064104}
  (\bibinfo{year}{2019}).

\bibitem{Qu-ACSAMI-2019}
\bibinfo{author}{Qu, J.}, \bibinfo{author}{Zhu, S.}, \bibinfo{author}{Zhang,
  W.} \& \bibinfo{author}{Zhu, Q.}
\newblock \bibinfo{journal}{\bibinfo{title}{Electrides with dinitrogen
  ligands}}.
\newblock {\emph{\JournalTitle{ACS Applied Materials \& Interfaces}}}
  \textbf{\bibinfo{volume}{11}}, \bibinfo{pages}{5256--5263},
  \doiprefix\url{10.1021/acsami.8b18676} (\bibinfo{year}{2019}).

\bibitem{NaBa2O-2001}
\bibinfo{author}{Vajenine, G.~V.} \& \bibinfo{author}{Simon, A.}
\newblock \bibinfo{journal}{\bibinfo{title}{{NaBa$_2$O}: A fresh perspective in
  suboxide chemistry}}.
\newblock {\emph{\JournalTitle{Angew. Chem. Int. Ed.}}}
  \textbf{\bibinfo{volume}{40}}, \bibinfo{pages}{4220--4222}
  (\bibinfo{year}{2001}).

\bibitem{Wan2018}
\bibinfo{author}{Wan, B.} \emph{et~al.}
\newblock \bibinfo{journal}{\bibinfo{title}{{Identifying quasi-2D and 1D
  electrides in yttrium and scandium chlorides via geometrical
  identification}}}.
\newblock {\emph{\JournalTitle{npj Computational Materials}}}
  \textbf{\bibinfo{volume}{4}}, \bibinfo{pages}{77},
  \doiprefix\url{10.1038/s41524-018-0136-1} (\bibinfo{year}{2018}).

\bibitem{NLO-1}
\bibinfo{author}{Muhammad, S.}, \bibinfo{author}{Xu, H.},
  \bibinfo{author}{Liao, Y.}, \bibinfo{author}{Kan, Y.} \& \bibinfo{author}{Su,
  Z.}
\newblock \bibinfo{journal}{\bibinfo{title}{Quantum mechanical design and
  structure of the li@b10h14 basket with a remarkably enhanced electro-optical
  response}}.
\newblock {\emph{\JournalTitle{J. Am. Chem. Soc.}}}
  \textbf{\bibinfo{volume}{131}}, \bibinfo{pages}{11833--11840},
  \doiprefix\url{10.1021/ja9032023} (\bibinfo{year}{2009}).

\bibitem{NLO-2}
\bibinfo{author}{Zhong, R.-L.}, \bibinfo{author}{Xu, H.-L.},
  \bibinfo{author}{Li, Z.-R.} \& \bibinfo{author}{Su, Z.-M.}
\newblock \bibinfo{journal}{\bibinfo{title}{Role of excess electrons in
  nonlinear optical response}}.
\newblock {\emph{\JournalTitle{The Journal of Physical Chemistry Letters}}}
  \textbf{\bibinfo{volume}{6}}, \bibinfo{pages}{612--619},
  \doiprefix\url{10.1021/jz502588x} (\bibinfo{year}{2015}).

\bibitem{Kim2006}
\bibinfo{author}{Kim, S.~W.}, \bibinfo{author}{Toda, Y.},
  \bibinfo{author}{Hayashi, K.}, \bibinfo{author}{Hirano, M.} \&
  \bibinfo{author}{Hosono, H.}
\newblock \bibinfo{journal}{\bibinfo{title}{{Synthesis of a Room Temperature
  Stable 12CaO{\textperiodcentered}7Al$_2$O$_3$ Electride from the Melt and Its
  Application as an Electron Field Emitter}}}.
\newblock {\emph{\JournalTitle{Chem. Mater.}}} \textbf{\bibinfo{volume}{18}},
  \bibinfo{pages}{1938--1944}, \doiprefix\url{10.1021/cm052367e}
  (\bibinfo{year}{2006}).

\bibitem{Vasp-PRB-1996}
\bibinfo{author}{Kresse, G.} \& \bibinfo{author}{Furthm\"uller, J.}
\newblock \bibinfo{journal}{\bibinfo{title}{Efficient iterative schemes for ab
  initio total-energy calculations using a plane-wave basis set}}.
\newblock {\emph{\JournalTitle{Phys. Rev. B}}} \textbf{\bibinfo{volume}{54}},
  \bibinfo{pages}{11169--11186}, \doiprefix\url{10.1103/PhysRevB.54.11169}
  (\bibinfo{year}{1996}).

\bibitem{PBE-PRL-1996}
\bibinfo{author}{Perdew, J.~P.}, \bibinfo{author}{Burke, K.} \&
  \bibinfo{author}{Ernzerhof, M.}
\newblock \bibinfo{journal}{\bibinfo{title}{Generalized gradient approximation
  made simple}}.
\newblock {\emph{\JournalTitle{Phys. Rev. Lett.}}}
  \textbf{\bibinfo{volume}{77}}, \bibinfo{pages}{3865--3868},
  \doiprefix\url{10.1103/PhysRevLett.77.3865} (\bibinfo{year}{1996}).

\bibitem{pymatgen-2013}
\bibinfo{author}{Ong, S.~P.} \emph{et~al.}
\newblock \bibinfo{journal}{\bibinfo{title}{Python materials genomics
  (pymatgen): A robust, open-source python library for materials analysis}}.
\newblock {\emph{\JournalTitle{Computational Materials Science}}}
  \textbf{\bibinfo{volume}{68}}, \bibinfo{pages}{314--319},
  \doiprefix\url{10.1016/j.commatsci.2012.10.028} (\bibinfo{year}{2013}).

\end{thebibliography}

\section*{SUPPLEMENTAL INFORMATION}
Supplemental Information can be found online at \url{https://www.cell.com/cms/10.1016/j.matt.2019.06.017/attachment/bd3df96a-d70e-4b2d-b702-fcf909e1711b/mmc1}

\section*{AUTHOR CONTRIBUTIONS}
Q.Z. designed the research. Q.Z., T.F., K.C. performed and analyzed the calculations and contributed to interpretation and discussion of the data. Q.Z. wrote the manuscript.


\section*{DECLARATION OF INTERESTS}
The authors declare no competing financial interests.

\end{document}